\documentclass[aps,prd,floats,floatfix,twocolumn,nofootinbib]{revtex4}
\usepackage{amsmath} 
\usepackage{amssymb} 
\usepackage{graphicx}
\usepackage{bm}

\newcommand{\half}{{\scriptstyle 1/2}}
\newcommand{\halfs}{{\scriptstyle\frac{1}{2}}}

\newcommand{\beq}{\begin{equation}}
\newcommand{\eeq}{\end{equation}}
\newcommand{\beqa}{\begin{eqnarray}}
\newcommand{\eeqa}{\end{eqnarray}}

\newcommand{\DeltaT}{\Delta T}

\begin{document}

\title{Optimal Constraint Projection for Hyperbolic Evolution
Systems}

\author{Michael Holst${}^{1,2}$, Lee Lindblom${}^1$, Robert Owen${}^1$,\\
Harald P. Pfeiffer${}^1$, Mark A. Scheel${}^1$, and Lawrence E. Kidder${}^3$
}
\affiliation{${}^1$ Theoretical Astrophysics 130-33, California Institute of
Technology, Pasadena, CA 91125}
\affiliation{${}^2$ Department of Mathematics, University of California
at San Diego, La Jolla, CA 92093}
\affiliation{${}^3$ Center for Radiophysics and Space Research, 
Cornell University, Ithaca, New York, 14853}

\date{\today}

\begin{abstract}
Techniques are developed for projecting the solutions of symmetric
hyperbolic evolution systems onto the constraint submanifold (the
constraint-satisfying subset of the dynamical field space). These
optimal projections map a field configuration to the ``nearest''
configuration in the constraint submanifold, where distances between
configurations are measured with the natural metric on the space of
dynamical fields.  The construction and use of these projections is
illustrated for a new representation of the scalar field equation that
exhibits both bulk and boundary generated constraint violations.
Numerical simulations on a black-hole background show that bulk
constraint violations cannot be controlled by constraint-preserving
boundary conditions alone, but are effectively controlled by
constraint projection.  Simulations also show that constraint
violations entering through boundaries cannot be controlled by
constraint projection alone, but are controlled by
constraint-preserving boundary conditions.  Numerical solutions to the
pathological scalar field system are shown to converge to solutions of
a standard representation of the scalar field equation when constraint
projection and constraint-preserving boundary conditions are used
together.
\end{abstract}


\maketitle
\section{Introduction}
\label{s:Introduction}

The exponential growth of constraint violations in the evolutions of
black-hole spacetimes is probably the most critical problem facing the
numerical relativity community today.  The evolution equations of any
self-consistent evolution system with constraints (including
Einstein's) ensure that if the constraints are satisfied
identically on an initial spacelike surface, they will remain satisfied
within the domain of dependence of that surface.  This does not mean
that small initial violations of the constraints will remain small, or
that constraint violations will not flow into the computational domain
through timelike boundaries.  On the contrary, experience has shown
that constraint violations seeded by roundoff or truncation level
errors in the initial data tend to grow exponentially in the numerical
evolutions of black-hole spacetimes (see {\it e.g.},~\cite{Kidder2001,
Lindblom2002, Scheel2002}). At present these constraint violating
instabilities are the limiting factor preventing these numerical
simulations from running for the desired length of time.  Finding ways
to control the growth of these constraints is therefore our most
urgent priority.

Recent work has demonstrated numerically that constraint violations
that flow into the computational domain through timelike boundaries
can be controlled effectively by the use of special constraint
preserving boundary conditions~\cite{Iriondo2002, Szilagyi2002,
Calabrese2002a, Szilagyi2003, Lindblom2004}.  A number of groups have
constructed such boundary conditions for various representations of
the Einstein evolution system~\cite{Stewart1998, FriedrichNagy1999,
Calabrese2002a, Szilagyi2002, Calabrese2003, Szilagyi2003,
Frittelli2003a, Frittelli2003b, Calabrese2003a, Frittelli2003c,
Alekseenko2004}.  However, constraint violations in many evolution
systems (including Einstein's) are driven by bulk terms in addition to
boundary terms in the constraint evolution equations.  In this paper
we demonstrate that such bulk generated constraint violations cannot
be controlled effectively through the use of boundary conditions
alone.  Alternative methods of controlling the growth of constraints
are still required in such systems.

The most widely used method of controlling the growth of constraints
in the Einstein evolution system is called fully constrained
evolution.  
In this method, which is often applied to spherical or axisymmetric
problems, symmetry considerations are used to separate the dynamical
fields into those that are determined by solving evolution equations
and those that are determined by enforcing the constraints at each
time step~\cite{StPi85, AbEv92, ch93, AbEv93, ACST94, Choptuik2003,
Choptuik2003a}.  In 3D problems without symmetry there is no obvious
way to perform such a separation in a general coordinate system; however,
fully constrained 3D methods based on spherical coordinates have
yielded promising results~\cite{Bonazzola2004}.
Various groups have studied a closely related method,
constraint projection, which can be used for general 3D evolutions in
any coordinate system.
The idea is to use the evolution system to advance all of the
dynamical fields in time, and then at each time step (or whenever the
constraints become too large) to force the solution back onto the
constraint submanifold by solving the
constraint equations (for the conformal factor and the longitudinal
part of the extrinsic curvature in the case of the Einstein system).
The first preliminary results obtained with this constraint projection
technique have been moderately successful~\cite{Anderson2003, Bonazzola2004,
Schnetter2003}.  Constraint projection has not gained widespread use
in 3D simulations, however, due in part to the traditionally high cost
of solving the elliptic constraint equations.  Difficult questions
also remain unanswered about the proper boundary conditions to impose
on the constraint equations, for example at black hole excision
boundaries.  Moreover, little attention has been given to the question
of whether these projections correctly map a field configuration onto
(or near) the correct point of the constraint submanifold, {\it i.e.}
the point through which the exact evolution of the system would pass
at that time.  In particular, it is not clear whether the overall time
evolution scheme---including the projections---remains consistent,
stable, and convergent.

The need to enforce constraints is a common feature of many problems
in mathematical physics besides numerical relativity, and for
many problems successful techniques have been developed to
ensure that numerical solutions satisfy the needed constraints.
Under mild assumptions on the constraints, the subset of
the field space satisfying the constraint equations defines a
formal differentiable manifold (a classical result due to
Ljusternik~\cite{Zeid91c}), and the evolution of a dynamical system
of ordinary (ODE) or partial differential equations (PDE) 
subject to constraints may be viewed as evolution on this submanifold.
Constraint control methods for such systems are generally based on ideas
from variational mechanics, where the Lagrangian (whose
stationary points describe the physical states of the system) is
augmented with a sum of terms consisting of products of Lagrange
multipliers and the constraints~\cite{Lanc49, Arno89, MaHu94}.
A necessary condition for a configuration point to be a solution of
both the field equations and the constraint equations is that the
augmented Lagrangian be stationary with respect to variations in both the
fundamental fields and the Lagrange multipliers~\cite{Zeid91c}.
The additional terms in the augmented Lagrangian involving Lagrange
multipliers can be viewed as forcing the dynamics to remain on the
constraint submanifold.

These augmented Lagrangian techniques are the basis of well-studied
numerical methods for controlling constraint violations in ODE
systems.  Many ODE systems are subject to algebraic constraints which
must be preserved as the solution evolves.  For such systems there
exist numerical integration techniques that enforce these algebraic
constraints exactly, and that also conserve various important
properties of the ODE solution (e.g.  time-reversibility and
symplectic structure).  These numerical techniques are derived by
adding to the ODEs terms chosen to make a suitable augmented
Lagrangian for the system stationary~\cite{Gear71, LeRe94, LeSk94,
Leim99, HLW01}. The resulting numerical schemes, referred to as
``step-and-project'' methods, can be thought of as standard time
integration steps followed by projections.  First a preliminary step
is taken forward in time using a standard numerical scheme, after
which the solution will generally not satisfy the constraint
equations.  Then the solution from the preliminary step is corrected
using a formal (optimal, or nearest point) projection back onto the
constraint submanifold.  This projection step typically involves
solving algebraic equations.  Unlike the simple constraint projection
methods used so far in numerical relativity, ``step-and-project''
numerical methods for constrained systems are well studied and well
understood.  It has been shown that they retain the consistency and
stability properties of the original one-step method on which they are
based, and they generally have the same convergence
properties~\cite{HLW01}.  These techniques are immediately applicable
to constrained PDE systems that are discretized in space to produce
constrained ODE systems (as we do, see Sec.~\ref{s:NumericalResults});
and numerical methods based on augmented Lagrangians for PDE systems
have also been developed~\cite{Tema77,BrFo91}.

In this paper we apply these augmented variational techniques to
obtain equations that project solutions of constrained hyperbolic
evolution systems onto the constraint submanifold of the appropriate
dynamical field space.  We construct projections that are optimal, in
the sense that they map a given field configuration to the ``nearest''
point on the constraint submanifold.  We use the natural metric, the
symmetrizer, that exists in any symmetric hyperbolic evolution system
to define distances on the space of fields.  Hence this optimal
projection is the one that minimizes this symmetrizer distance
(typically called the energy) between a given field configuration and
its projection.  The general formalism for constructing such
optimal projections for constrained hyperbolic evolution systems is
described in Sec.~\ref{s:OptimalProjection}.

We illustrate these optimal constraint projection ideas in
Sec.~\ref{s:ScalarFields} by deriving the optimal projection for a new
symmetric-hyperbolic representation of the scalar field equation on a
fixed background spacetime.  This scalar field system has the
interesting property that it suffers from constraint violations driven
both by bulk terms as well as boundary flux terms in the equations.
(And so this system serves as a good model of the pathologies present
in the Einstein system.)  The optimal projection for this scalar field
system is determined by solving a certain elliptic PDE.  In
Sec.~\ref{s:NumericalResults} we test these optimal projection
techniques by studying numerical solutions to this scalar field system
on a fixed black-hole background spacetime.  In particular we
demonstrate that constraint preserving boundary conditions are
necessary, but not sufficient, to control the growth of constraints in
this pathological scalar field system.  We demonstrate that constraint
projection succeeds in producing convergent constraint-satisfying
solutions, but only if constraint preserving boundary conditions are
used as well.  These tests also illustrate that the projections are
best performed at fixed time intervals ($\DeltaT\approx 2M$ for this
problem) rather than after each time step.  And we show that the
computational cost of solving the constraint projection equations for
this system (using our spectral elliptic solver~\cite{Pfeiffer2003})
is a very small fraction (below $1\%$ for the resolution needed to
achieve roundoff level accuracy) of the total computational cost of
evolving this system.  The symmetrizer metric for this model scalar
field system (like many hyperbolic evolution systems) is not unique;
so the projections defined in terms of the symmetrizer are not unique.
Nevertheless, we demonstrate for the model scalar field system that
numerical evolutions based on these different projections all converge
to the same solution.  The rate of this convergence is not the same
for all projections, however, and we find an ``optimal'' projection
for this system that maximizes this convergence rate.

\section{Optimal Constraint Projection}
\label{s:OptimalProjection}

Our objective is to construct a projection operator that maps a given
field configuration to the nearest constraint-satisfying
configuration (the nearest point on the constraint submanifold).
That is, we wish to map an initial point $\bar{u}^\alpha$ in the
field configuration space to a new point $u^\alpha$ that satisfies a
set of constraint equations:
\begin{eqnarray}
c^A(u^\alpha)=0.\label{e:constrainteq}
\end{eqnarray}
(We use Greek indices to label individual components of the
dynamical fields, and upper case Latin indices to label the
individual components of the constraints.)  To find the optimal
projection we also need to have a distance measure between field
points. We define the needed measure in terms of a symmetric
positive-definite metric, $S_{\alpha\beta}$, on the dynamical field
space.  The distance between field points is then defined as
\begin{eqnarray}
||\,\delta u\,||^2 = \int S_{\alpha\beta}(u^\alpha-\bar u^\alpha)
(u^\beta-\bar u^\beta)d^{\,3}x.
\end{eqnarray}
Building on the augmented variational techniques commonly used to
construct step-and-project constraint control schemes
in other areas of numerical
analysis~\cite{Gear71,LeSk94,HLW01},
we are now prepared to construct the optimal
projection map.  We introduce a Lagrangian density $\cal L$ that consists of
the distance between the given field configuration $\bar u^\alpha$ and
its projection $u^\alpha$, plus the products of the constraints with
Lagrange multipliers.  Thus we introduce the Lagrangian density,
\begin{eqnarray}
{\cal L} = S_{\alpha\beta}(u^\alpha-\bar u^\alpha)(u^\beta-\bar u^\beta)
+\lambda_A c^A.\label{e:lagrangian0}
\end{eqnarray}
The stationarity of the Lagrangian (the volume integral of this
Lagrangian density) with respect to variations of the Lagrange
multipliers $\lambda_A$ enforces the constraints, while stationarity
with respect to variations of the fields $u^\alpha$ are necessary
conditions for the projection to minimize the distance to the
constraint submanifold.

The optimal projection procedure outlined above could be carried out
using any definition of the distance between field points, {\it e.g.}
using any positive definite metric $S_{\alpha\beta}$ on the space of
fields.  For a particular problem this distance measure should be
chosen to be the natural measure associated with that problem.  Our
primary interest here is the construction of projections for
constrained hyperbolic evolution systems.  So we will focus our
attention on fields $u^\alpha$ that satisfy a first-order evolution
equation of the form
\begin{eqnarray}
\partial_t u^\alpha+A^{k\alpha}{}_\beta\partial_ku^\beta=
F^\alpha.
\end{eqnarray}
We use lower case Latin indices to label spatial coordinates $x^k$,
$\partial_t=\partial/\partial t$ to denote time derivatives, and
$\partial_k=\partial/\partial x^k$ to denote spatial derivatives.
Such systems are called symmetric hyperbolic if they have a positive
definite metric $S_{\alpha\beta}$ on the space of fields (typically
called the symmetrizer) that symmetrizes the characteristic matrices:
\begin{eqnarray}
S_{\alpha\gamma}A^{k\gamma}{}_\beta\equiv A^k_{\alpha\beta}= A^k_{\beta\alpha}.
\end{eqnarray}
The well-posedness of the initial value problem for linear
symmetric-hyperbolic evolution systems is demonstrated by establishing
bounds on the square-integral norm of the dynamical fields defined
with this symmetrizer metric~\cite{Tayl96c,Gustafsson1995}.  This
metric defines the meaningful measure on the dynamical field space for
symmetric-hyperbolic systems, so this is the appropriate measure to
use for constructing optimal constraint projections for these systems.
Most hyperbolic evolution systems of interest in mathematical physics
(including many representations of the Einstein system) are symmetric
hyperbolic, and so we limit our consideration here to systems of this
type.

In Sec.~\ref{s:ScalarFields} we use the procedure outlined above to
construct explicitly the optimal projection for the relatively simple
case of the scalar wave equation on a curved background spacetime.
But before we focus on that special case, we take a few (rather more
abstract) steps in the construction of this projection for the general
case.  To do this we assume that the constraints $c^A$ are linear
in the derivatives of the dynamical fields:
\begin{eqnarray}
c^A=K^{kA}{}_\beta \partial_k u^\beta + L^A,
\end{eqnarray}
where $K^{kA}{}_\beta$ and $L^A$ may depend on $u^\alpha$ but not its
derivatives.  The constraints have this general form in many
evolution systems of interest ({\it e.g.,} the Einstein
system, the Maxwell system, the incompressible fluid system).  
In this case we can explicitly compute the variations of the
Lagrangian density defined in Eq.~\eqref{e:lagrangian0}:
\begin{eqnarray}
&\frac{\delta{\cal L}}{\delta u^\alpha}\delta u^\alpha   &=
\delta u^\alpha\Bigl\{2S_{\alpha\beta}
(u^\beta-\bar u^\beta)-\partial_k(\lambda_A K^{kA}{}_\alpha)\nonumber\\
&&\quad+\lambda_A\bigl(\partial_\alpha K^{kA}{}_\beta\partial_k u^\beta
+\partial_\alpha L^A\bigr)\Bigr\}
\,\nonumber\\
&&\quad+\partial_k\bigl(\lambda_A K^{kA}{}_\alpha\delta u^\alpha\bigr),\\
&\frac{\delta{\cal L}}{\delta \lambda_A}\delta \lambda_A &= c^A\delta\lambda_A.
\end{eqnarray}
Here we use the notation $\partial_\alpha \equiv \partial/\partial
u^\alpha$ to denote derivatives with respect to the fields.  We have
also assumed that the symmetrizer $S_{\alpha\beta}$ may depend on
$\bar u^\alpha$ but not $u^\alpha$.  We wish to find the stationary
points of this Lagrangian with respect
to arbitrary variations in the fields $u^\alpha$ and the Lagrange
multipliers $\lambda_A$.  Stationarity with respect to the variations
of these quantities (that vanish on the boundaries) implies that
\begin{eqnarray}
0&=&\!u^\alpha-\bar
u^\alpha-\halfs S^{\alpha\beta}\partial_k(\lambda_A K^{kA}{}_\beta)\nonumber\\
&&\qquad+\halfs\lambda_AS^{\alpha\beta}
\bigl(\partial_\beta K^{kA}{}_\gamma\partial_k
u^\gamma +\partial_\beta L^A\bigr),\label{e:varu}\\
0&=&c^A=K^{kA}{}_\beta \partial_k u^\beta + L^A\label{e:varlam}
\end{eqnarray}
at each interior point, and stationarity with respect to boundary
variations implies that
\begin{eqnarray}
0=n_k \lambda_A K^{kA}{}_\beta\label{e:boundary0}
\end{eqnarray}
at each boundary point, where $n_k$ is the outward directed unit
normal to the surface.  We use the notation $S^{\alpha\beta}$ to
denote the inverse of $S_{\alpha\beta}$.  The general idea is to use
Eqs.~\eqref{e:varu} and \eqref{e:varlam}, with appropriate boundary
conditions (such as those provided by Eq.~[\ref{e:boundary0}]), to
determine the field configuration $u^\beta$ and the Lagrange
multipliers $\lambda_A$ for any given field point $\bar u^\alpha$.  If
$u^\alpha$ and $\lambda_A$ satisfying these equations can be found,
then we are guaranteed that the field $u^\alpha$ is the
constraint-satisfying solution nearest the point $\bar u^\alpha$ as
desired.  We do not know whether these equations always admit
solutions in the general case.  So in Sec.~\ref{s:ScalarFields} we
study in detail this optimal projection for the simple case of the
scalar field equations on a fixed background spacetime.  We show that
solutions to the optimal projection equations always exist and are
relatively easy to compute numerically in this simple case.  And in
Sec.~\ref{s:NumericalResults} we show that this optimal projection is
very effective in controlling the growth of constraints for the scalar
field system.

\section{Scalar Fields in Curved Spacetime}
\label{s:ScalarFields}
In this section we examine in some detail the scalar wave system on a
fixed background spacetime.  In Sec.~\ref{s:ModifiedScalarWaveSystem}
we review the standard treatment of this system, and then modify it so
that it exhibits bulk generated constraint violations in addition to
the boundary generated violations present in the standard system.
This new, more pathological, symmetric-hyperbolic scalar field system
now serves as a good model of the constraint violating problems
inherent in the Einstein system.  We construct constraint
preserving boundary conditions for this system in
Sec.~\ref{s:ConstraintPreservingBoundaryConditions}, and the optimal
projection map for this system in
Sec.~\ref{s:OptimalConstraintProjection} following the procedure
outlined in Sec.~\ref{s:OptimalProjection}.

\subsection{Modified Scalar Wave System}
\label{s:ModifiedScalarWaveSystem}
The standard scalar field equation on a fixed background spacetime is
\begin{equation}
\nabla^\mu\nabla_\mu\psi = 0,\label{e:waveeq}
\end{equation}
where $\psi$ represents the scalar field and $\nabla_\mu$ the
covariant derivative associated with the background spacetime metric.  We
represent the background spacetime metric in terms of the usual $3+1$
splitting:
\begin{equation}
ds^2=-N^2 dt^2+g_{ij}(dx^i+N^i dt)(dx^j+N^jdt),
\end{equation}
where the lapse $N$ and the spatial metric $g_{ij}$ are assumed to be
positive definite, while the shift $N^i$ is arbitrary.  The equation
for the scalar field $\psi$, Eq.~\eqref{e:waveeq}, can be re-expressed
as a first-order evolution system in the standard way (see {\it e.g.}
Ref.~\cite{Scheel2004}):
\begin{eqnarray}
&\partial_t\psi - N^k\partial_k\psi &= -N\Pi,\label{e:psidot}\\
&\partial_t\Pi  - N^k\partial_k\Pi  +Ng^{ki}\partial_k\Phi_i
&=  N J^i\Phi_i + NK\Pi,\label{e:pidot}\\
&\partial_t\Phi_i - N^k\partial_k\Phi_i+N\partial_i\Pi
&= -\Pi\partial_iN+\Phi_j\partial_iN^j.\qquad\label{e:Phidot}
\end{eqnarray}
The field $\Phi_i$ represents the spatial gradient $\partial_i\psi$,
and $\Pi$ represents the time derivative of $\psi$ (and is defined
precisely by Eq.~[\ref{e:psidot}]).  The auxiliary quantities $K$ (the
trace of the extrinsic curvature) and $J^i$ depend only on the
background spacetime geometry, and are defined by
\begin{eqnarray}
J^i&=&-N^{-1}g^{-\half}\partial_j(N g^\half g^{ij}),\\
K&=&-N^{-1}g^{-\half}\bigl[\partial_t g^{\half}
-\partial_j(g^{\half}N^j)\bigr].
\end{eqnarray}

Solutions to the first-order system,
Eqs.~\eqref{e:psidot}--\eqref{e:Phidot}, are also solutions to
Eq.~\eqref{e:waveeq} only if the constraints are satisfied:
$0=c^A\equiv\{{\cal C}_i,{\cal C}_{ij}\}$, where
\begin{eqnarray}
{\cal C}_i&=& \partial_i\psi - \Phi_i,\label{e:c1def}\\
{\cal C}_{ij}&=&\partial_{[i}\Phi_{j]}.\label{e:c2def}
\end{eqnarray}
Although the second constraint, $C_{ij}=0$, follows from the first,
$C_{i}=0$, the converse is not true. Hence we include both constraints
in the analysis here.  Note that both constraints are necessary to
construct a first-order hyperbolic evolution system for the constraint
quantities (discussed below, Eqs.~(\ref{e:const1l})
and~(\ref{e:const2l})).  Note also that the analogues of both
constraints play essential roles in first-order hyperbolic
formulations of Einstein's equations.

We now generalize the evolution system,
Eqs.~\eqref{e:psidot}--\eqref{e:Phidot}, somewhat by adding multiples
of the constraint ${\cal C}_i$ to Eqs.~\eqref{e:psidot} and
\eqref{e:Phidot}:
\begin{eqnarray}
&\partial_t\psi - N^k\partial_k\psi &= -N\Pi+\gamma_1 N^k{\cal C}_k,\qquad
\label{e:psidottemp}\\
&\partial_t\Phi_i - N^k\partial_k\Phi_i+N\partial_i\Pi
&= -\Pi\partial_iN+\Phi_j\partial_iN^j\nonumber\\
&&\quad\,+\gamma_2N{\cal C}_i,\label{e:phidottemp}
\end{eqnarray}
where $\gamma_1$ and $\gamma_2$ are arbitrary constants.  The
constraint-satisfying solutions to these equations are the same as
those of the original system; but as we shall see, the constraint
violating properties of the new system are significantly different
from those of the original.  Substituting the definition of ${\cal
C}_i$ in Eqs.~\eqref{e:psidottemp} and \eqref{e:phidottemp} gives us new
evolution equations for $\psi$ and $\Phi_i$:
\begin{align}
\partial_t\psi& - (1+\gamma_1)N^k\partial_k\psi = -N\Pi-\gamma_1 N^k\Phi_k,
\label{e:psidotnew}\\
\partial_t\Phi_i& - N^k\partial_k\Phi_i + N\partial_i\Pi 
- \gamma_2N\partial_i\psi
=\!\nonumber\\
&\qquad\qquad\qquad\,\,\,
-\Pi\partial_iN+\Phi_j\partial_iN^j
-\gamma_2N\Phi_i.\label{e:Phidotnew}
\end{align}

The first-order system that represents the scalar wave equation,
Eqs.~\eqref{e:pidot}, \eqref{e:psidotnew}, and \eqref{e:Phidotnew}, has
the standard first order form,
\begin{eqnarray}
\partial_tu^\alpha+A^{k\alpha}{}_\beta\partial_ku^\beta&=&F^\alpha,
\label{e:evolsystem}
\end{eqnarray}
where $u^\alpha=\{\psi,\Pi,\Phi_i\}$.  Systems of this type are called
symmetric hyperbolic if there exists a symmetric positive-definite
tensor $S_{\alpha\beta}$ on the space of fields that symmetrizes the
characteristic matrices $A^{k\alpha}{}_\beta$:
\begin{equation}
S_{\alpha\gamma}A^{k\gamma}{}_\beta\equiv A^k_{\alpha\beta}=A^k_{\beta\alpha}.
\label{eq:Symmetrizer}
\end{equation}  
The most general symmetrizer for our new scalar wave system is (up to
an overall factor),
\begin{eqnarray}
ds^2&=&S_{\alpha\beta}du^\alpha du^\beta,\nonumber\\
&=&\Lambda^2 d\psi^2 -2\gamma_2 d\psi\, d\kern 1pt\Pi
+  d\kern 1pt\Pi^2 + g^{ij}d\Phi_id\Phi_j,\qquad
\label{eq:ScalarWaveSymmetrizer}
\end{eqnarray}
where $\Lambda$ is an arbitrary non-vanishing function.  This
$S_{\alpha\beta}$ symmetrizes the characteristic matrices
$A^k_{\alpha\beta}$ so long as $\gamma_1\gamma_2=0$.  Thus we must
take at least one of these parameters to be zero for our new system to
be symmetric hyperbolic.  This symmetrizer is positive definite whenever
\begin{equation}\label{eq:gamma2-condition}
\Lambda^2>\gamma_2^2.
\end{equation}
In this case $S_{\alpha\beta}$ provides a dynamically meaningful
measure of the distance between field configurations, which we use to
define our optimal constraint projection operator in
Sec.~\ref{s:OptimalConstraintProjection}.

The evolution of the constraints follows from the principal evolution
system, Eqs.~\eqref{e:pidot}, \eqref{e:psidotnew}, and~\eqref{e:Phidotnew}:
\begin{eqnarray}
&\partial_t{\cal C}_{i}-(1+\gamma_1){\cal L}_{\vec N}{\cal C}_{i}&
=2\gamma_1 N^j{\cal C}_{ji}-\gamma_2N{\cal C}_i,\label{e:const1l}\\
&\partial_t{\cal C}_{ij}-{\cal L}_{\vec N}{\cal C}_{ij}&=
-\gamma_2N{\cal C}_{ij}-\gamma_2{\cal C}_{[i}\partial_{j]}N,\qquad
\label{e:const2l}
\end{eqnarray}
where ${\cal L}_{\vec N}$ represents the Lie derivative along the
shift vector $N^i$.  If the constraints are satisfied at some initial
time, then these equations guarantee (at least at the analytical
level) that the constraints remain satisfied in the domain of
dependence of the initial data.  These equations also show that any
constraint violations in this system will be advected along a
congruence of timelike curves.  Constraint violations can therefore
flow into the computational domain if these curves intersect the
boundaries.  And like the Einstein evolution system, these equations
also contain bulk terms that amplify any existing constraint
violations.  When $\gamma_1=0$ we see that Eq.~\eqref{e:const1l}
implies that the constraint ${\cal C}_i$ has the simple time
dependence ${\cal C}_i(\tau)={\cal C}_i(0) e^{-\gamma_2\tau}$, where
$\tau$ measures proper time as seen by a hypersurface orthogonal
observer.  Whenever $\gamma_2<0$ this constraint grows exponentially,
and in this case the modified scalar wave system serves as a good
model of the constraint violations in the Einstein system.
(Constraint violations of all wavelengths grow exponentially in this
system, and so it may be even more pathological than the Einstein
system where constraint violating instabilities are typically
dominated by long wavelength modes~\cite{Lindblom2002,Scheel2002}.)
Conversely, if $\gamma_2>0$ then this modified scalar wave system
exponentially suppresses any residual constraint violations that may
be present in the initial data.  This latter property suggests that
analogous terms could be introduced to control some of the bulk
constraint violating terms in the Einstein system.
  
\subsection{Constraint Preserving Boundary Conditions}
\label{s:ConstraintPreservingBoundaryConditions}

Boundary conditions for hyperbolic evolution systems are defined in
terms of the characteristic fields of these systems, so we must
construct these fields for our modified scalar wave system.  The
characteristic fields are defined with respect to a spatial direction
at each point, represented here by a unit normal co-vector field
$n_k$.  For the purposes of imposing boundary conditions, the
appropriate $n_k$ is the outward-pointing normal to the boundary.
Given a direction field $n_k$ we define the left eigenvectors
$e^{\hat \alpha}{}_\alpha$ of the characteristic matrix
$n_kA^{k\,\alpha}{}_\beta$ by
\begin{eqnarray}
e^{\hat\alpha}{}_\alpha n_k A^{k\,\alpha}{}_\beta=v_{(\hat\alpha)} 
e^{\hat\alpha}{}_\beta,\label{e:eigenvalueeq}
\end{eqnarray}
where $v_{(\hat \alpha)}$ denotes the eigenvalue (also called the
characteristic speed).  The index $\hat \alpha$ labels the various
eigenvectors and eigenvalues, and there is no summation over
$\hat\alpha$ in Eq.~\eqref{e:eigenvalueeq}.  Since we are interested
in hyperbolic evolution systems, the space of eigenvectors has
the same dimension as the space of dynamical fields, and the matrix
$e^{\hat\alpha}{}_\beta$ is invertible.  The projections of the
dynamical fields $u^\alpha$ onto these left eigenvectors are
called the characteristic fields $u^{\hat\alpha}$:
\begin{eqnarray}
u^{\hat\alpha} = e^{\hat\alpha}{}_{\beta}u^\beta.
\label{e:characteristicparts}
\end{eqnarray}
At each boundary point, boundary conditions must be imposed on any
characteristic field having
negative characteristic speed, $v_{(\hat\alpha)}<0$, at 
that point~\cite{Kevo90,Tayl96a}.
We refer to fields with $v_{(\hat\alpha)}<0$ as the incoming
characteristic fields at that point.  Conversely, those characteristic
fields having non-negative characteristic speeds (the outgoing fields)
must not have boundary conditions imposed on them there.

The characteristic fields for the symmetric hyperbolic representations
($\gamma_1 \gamma_2 = 0$) of the scalar wave system are the
quantities $u^{\hat\alpha}=\{Z^1,Z^2_i,U^{1\pm}\}$:
\begin{eqnarray}
Z^1&=&\psi,\label{eq:Z1definition}\\
Z^2_i&=&P^k{}_i\Phi_k,\\
U^{1\pm}&=&\Pi\pm n^k\Phi_k-\gamma_2\psi,\label{eq:U1dfinition}
\end{eqnarray}
where $P^k{}_i=\delta^k{}_i-n^kn_i$, $n^k=g^{kj}n_j$, and $n^k n_k =
1$.  The fundamental fields $u^\alpha$ can be reconstructed from the
characteristic fields $u^{\hat\alpha}$ by inverting
Eq.~(\ref{e:characteristicparts}):
\begin{eqnarray}
\psi&=&Z^1,\\
\Pi&=&\halfs(U^{1+}+U^{1-})+\gamma_2 Z^1,\\
\Phi_i&=&\halfs(U^{1+}-U^{1-})n_i+Z^2_i\label{eq:PhiCharDecomp}.
\end{eqnarray}
The characteristic
field $Z^1$ propagates with speed $-\gamma_1 n_kN^k/N$, the field
$Z^2_i$ with speed 0, and the fields $U^{1\pm}$ with speeds $\pm 1$
relative to the hypersurface orthogonal
observers.  The coordinate characteristic speeds of these fields are
$-(1+\gamma_1)n_kN^k$, $-n_kN^k$ and $-n_kN^k\pm N$ respectively.

At each boundary point, boundary conditions are required on
each characteristic field whose coordinate characteristic speed is negative
at that point.
The field $U^{1-}$, in particular, requires a boundary condition on
all timelike boundaries.  For the standard representation of the
scalar field system, Eqs.~(\ref{e:psidot})--(\ref{e:Phidot}),
the boundary condition $U^{1-}=\Pi-n^k\Phi_k=0$
is used to ensure (approximately) 
that no scalar waves enter the computational domain.
We wish to enforce this condition on our generalized scalar field
system, Eqs.~(\ref{e:pidot}), (\ref{e:psidotnew}), and~(\ref{e:Phidotnew}),
in such a way that the physical (constraint satisfying)
solutions are the same for all values of the parameters $\gamma_1$ and
$\gamma_2$.  Since $U^{1-}$ depends on $\gamma_2$,
Eq.~\eqref{eq:U1dfinition}, the proper boundary condition must also
depend on $\gamma_2$: $U^{1-}+\gamma_2\psi=\Pi-n^k\Phi_k=0$.  Thus the
appropriate boundary condition to impose on $U^{1-}$ is
$U^{1-}=-\gamma_2\psi$.  The freezing form of this boundary
condition (as used in our code) is,
\begin{eqnarray}
\partial_t U^{1-}=-\gamma_2\partial_t\psi.
\label{eq:FreezingBcUm}
\end{eqnarray}

For boundary conditions on the fields $Z^1$ and $Z^2_i$ (when
necessary), we explore two choices: One is the freezing boundary
condition $\partial_tZ^i=\partial_tZ^2_i=0$.  In
Sec.~\ref{s:NumericalResults} we show that this boundary condition
allows constraint violations to enter the computational domain through
the boundaries.  Therefore, we also explore conditions that prevent
this influx of constraint violations: When the fields $Z^1$ and/or
$Z^2_i$ require boundary conditions, we set
\begin{eqnarray}
\partial_t Z^1&=& N^k\Phi_k -N\Pi,
\label{eq:ConstraintBcZ1}\\
\partial_t Z^2_i&=& P^k{}_i\partial_k\partial_t\psi.
\label{eq:ConstraintBcZ2}
\end{eqnarray}
Equation~(\ref{eq:ConstraintBcZ1}) is based on Eq.~(\ref{e:psidot})
combined with Eq.~(\ref{e:c1def}), while Eq.~(\ref{eq:ConstraintBcZ2})
is derived from the time-derivative of Eq.~(\ref{e:c1def}).  We note
that with the choice $\gamma_1=-1$, the field $Z^1$ never requires a
boundary condition.  We also note that the term $\partial_t\psi$ that
appears on right side of Eqs.~(\ref{eq:FreezingBcUm}) and
(\ref{eq:ConstraintBcZ2}) must be evaluated using the appropriate
expression for $\partial_t\psi=\partial_tZ^1$ on this boundary:
Eq.~(\ref{eq:ConstraintBcZ1}) when $Z^1$ requires a boundary
condition, or Eq.~(\ref{e:psidotnew}) when no boundary condition is
required.  In Sec.~\ref{s:NumericalResults} we compare numerically the
results of using these constraint preserving boundary conditions with
the use of the freezing boundary conditions $\partial_t Z^1=\partial_t
Z^2_i=0$ on these fields.

\subsection{Optimal Constraint Projection}
\label{s:OptimalConstraintProjection}

The idea is to use the full evolution system, Eqs.~\eqref{e:pidot},
\eqref{e:psidotnew}, and \eqref{e:Phidotnew}, to evolve initial data
forward in time an amount $\DeltaT$ and then (when the constraint
violations become too large) to project this solution back onto the
constraint submanifold in some optimal way.  Let $\bar
u^\alpha=\{\bar\psi,\bar\Pi, \bar\Phi_i\}$ denote the solution
obtained directly from this free evolution step.  This solution $\bar
u^\alpha$ may not satisfy the constraints because 
roundoff or truncation level constraint violations have been
amplified, or constraint violations have flowed through the boundaries.  
Thus we wish to project $\bar
u^\alpha$ in an optimal way back onto the constraint
submanifold.  Following the procedure outlined in
Sec.~\ref{s:OptimalProjection} we construct a Lagrangian density,
\begin{eqnarray}
{\cal L}&=&g^\half\bigl[
S_{\alpha\beta}(u^\alpha-\bar u^\alpha)(u^\beta - \bar u^\beta)
+\lambda_A c^A\bigr]\nonumber\\
&=&g^\half\Bigl[
\Lambda^2 (\psi-\bar\psi)^2-2\gamma_2(\psi-\bar\psi)(\Pi-\bar\Pi)\nonumber\\
&&\quad\;\; +(\Pi-\bar\Pi)^2 +g^{ij}(\Phi_i-\bar\Phi_i)(\Phi_j-\bar\Phi_j)
\nonumber\\
&&\quad\;\;+\lambda^i(\partial_i\psi-\Phi_i)
+\lambda^{ij}\partial_{[i}\Phi_{j]}\Bigr],\quad\label{e:lagrangian}
\end{eqnarray}
using the symmetrizer $S_{\alpha\beta}$ of the hyperbolic evolution system,
Eq.~(\ref{eq:ScalarWaveSymmetrizer}),
and the Lagrange multipliers $\lambda_A=\{\lambda^i,\lambda^{ij}\}$.
The stationary points of the Lagrangian,
\begin{eqnarray}
 L=\int{\cal L}\,d^{\,3}x,\label{e:lagrangian1}
\end{eqnarray}
with respect to variations in
$u^\alpha$ and $\lambda_A$ define the
optimally projected field configuration $u^\alpha$.  We have included
the multiplicative factor $g^\half=(\det\, g_{ij})^\half$ in
Eq.~\eqref{e:lagrangian} to ensure that $L$ is coordinate invariant.  

The scalar field constraint Lagrangian density,
Eq.~\eqref{e:lagrangian}, has the following variations:
\begin{eqnarray}
&\frac{\delta{\cal L}}{\delta\psi}\delta\psi &=
2g^\half \bigl[\Lambda^2(\psi-\bar\psi)-\gamma_2(\Pi-\bar\Pi)\bigr]\delta\psi
\nonumber\\
&&\quad-\partial_i(g^\half\lambda^i)\delta\psi
+\partial_i(g^\half\lambda^i\delta\psi),\label{eq:dL/dpsi}\\
&\frac{\delta{\cal L}}{\delta\Pi}\delta\Pi &=2g^\half
\bigl[\Pi-\bar\Pi-\gamma_2(\psi-\bar\psi)\bigr]\delta\Pi,\\
&\frac{\delta{\cal L}}{\delta\Phi_i}\delta\Phi_i &=
\bigl[2g^\half g^{ij}(\Phi_i-\bar\Phi_i)-g^\half\lambda^j\nonumber\\
&&\quad-\partial_i(g^\half\lambda^{ij})\bigr]\delta\Phi_j
+\partial_i(g^\half\lambda^{ij}\delta\Phi_j),\label{eq:dL/dPhi}\\
&\frac{\delta{\cal L}}{\delta\lambda^i}\delta\lambda^i &=
g^\half(\partial_i\psi-\Phi_i)\delta\lambda^i,\label{eq:dL/dlambda1}\\
&\frac{\delta{\cal L}}{\delta\lambda^{ij}}\delta\lambda^{ij} &=
g^\half\partial_{[i}\Phi_{j]}\delta\lambda^{ij}.\label{eq:dL/dlambda2}
\end{eqnarray}
We require that the Lagrangian $L$ from Eq.~\eqref{e:lagrangian1}
be stationary with respect to
{\em all} variations in the dynamical fields $\delta
u^\alpha=\{\delta\psi, \delta\Pi,\delta\Phi_i\}$ (including those
that do not vanish on the boundaries) as well as all variations in
the Lagrange multipliers $\delta\lambda_A
=\{\delta\lambda^i,\delta\lambda^{ij}\}$.  From
Eqs.~(\ref{eq:dL/dpsi})--(\ref{eq:dL/dPhi}), it follows that
\begin{eqnarray}
\psi&=&\bar\psi+\gamma_2\Lambda^{-2}(\Pi-\bar\Pi)
+\halfs\Lambda^{-2}g^{-\half}\partial_i(g^\half \lambda^i),\qquad
\label{e:psibar}\\
\Pi&=&\bar\Pi+\gamma_2(\psi-\bar\psi),\label{e:pibar}\\
\Phi_i&=&\bar\Phi_i+\halfs g_{ij}\lambda^j
+\halfs g^{-\half}g_{ij}\partial_k(g^\half\lambda^{kj}),\label{e:Psibar}
\end{eqnarray}
and Eqs.~(\ref{eq:dL/dlambda1}) and~(\ref{eq:dL/dlambda2}) imply
that the projected solution satisfies the constraints.  We now
solve Eq.~\eqref{e:Psibar} for $\lambda^i$, substitute it into
Eq.~\eqref{e:psibar}, and use Eqs.~(\ref{eq:dL/dlambda1})
and~\eqref{e:pibar}, to obtain the following equation for $\psi$,
\begin{eqnarray}
\nabla^i\nabla_i\psi-(\Lambda^2-\gamma_2^2)\psi &=&
\nabla^i\bar\Phi_i-(\Lambda^2-\gamma_2^2) \bar\psi,\label{e:helmholtz}
\end{eqnarray}
where $\nabla_i$ represents the spatial covariant derivative that is
compatible with $g_{ij}$.  In deriving this equation we have also used
the fact that the term $\partial_i\partial_k(g^\half \lambda^{ki})$
vanishes identically because $\lambda^{ij}$ is antisymmetric.
Equation~\eqref{e:helmholtz} is just the covariant inhomogeneous
Helmholtz equation. We note that the parameters must satisfy the
condition $\Lambda^2-\gamma_2^2>0$ for the evolution system to be
symmetric hyperbolic.  Solving Eq.~\eqref{e:helmholtz} determines the
optimal projection $\psi$; the optimal $\Pi$ is determined from
Eq.~\eqref{e:pibar},
\begin{equation}\label{eq:Pi-projection}
\Pi=\bar\Pi+\gamma_2(\psi-\bar\psi);
\end{equation}
and the optimal $\Phi_i$ is obtained by enforcing the constraint,
\begin{equation}\label{eq:Phi-Projection}
\Phi_i=\partial_i\psi.
\end{equation}
We note that the Lagrange multiplier
$\lambda^{ij}$ does not play any essential role in this analysis: we
could just as well have set $\lambda^{ij}=0$ and still obtained the
same projection.  This makes sense, because the constraint ${\cal
C}_{ij}$ is really a consequence of the constraint ${\cal C}_i$ in 
this case.

The evolution equations for $\Pi$ and $\Phi_i$, Eqs.~\eqref{e:pidot}
and \eqref{e:Phidot}, decouple from the larger scalar field evolution
system, Eqs.~\eqref{e:pidot}, \eqref{e:psidotnew}, and
\eqref{e:Phidotnew}, when $\gamma_2=0$.  It is sometimes of interest to
consider the properties of this smaller system, Eqs.~\eqref{e:pidot}
and \eqref{e:Phidot}, subject to the single
constraint, Eq.~\eqref{e:c2def}.  The optimal constraint projection
for this reduced system consists of Eqs.~\eqref{e:pibar} and
\eqref{e:Psibar} (with $\lambda^i=\gamma_2=0$), together with the
single constraint equation $\partial_{[i}\Phi_{j]}=0$.  This
constraint equation implies that $\Phi_i=\partial_i\psi$ for some
scalar function $\psi$.  Inserting this expression for $\Phi_i$ in
Eq.~\eqref{e:Psibar}, multiplying by $g^\half g^{ij}$, and taking the
divergence, we obtain the following equation for $\psi$,
\begin{eqnarray}
\nabla^i\nabla_i\psi&=&
\nabla^i\bar\Phi_i.\label{e:poisson}
\end{eqnarray}
In deriving this equation we have used the fact that the term
$\partial_i\partial_k(g^\half \lambda^{ki})$ vanishes identically
because $\lambda^{ij}$ is antisymmetric.  The optimal projection in
this reduced system then sets $\Pi=\bar\Pi$ and
$\Phi_i=\partial_i\psi$, where $\psi$ is the solution to
Eq.~\eqref{e:poisson}.  We note that Eq.~(\ref{e:poisson}) is just the
$\Lambda^2-\gamma_2^2=0$ limit of the original projection
Eq.~\eqref{e:helmholtz}.

Unfortunately the optimal constraint projection for the scalar field
system is not unique, because the parameter $\Lambda$ in the
symmetrizer metric is not unique.  We have seen that taking the limit
$\Lambda^2\rightarrow \gamma_2^2$ is equivalent to ignoring the
evolution of the scalar field $\bar\psi$ in constructing the optimal
projection.  Alternatively, the limit $\Lambda\rightarrow\infty$
corresponds to the simple projection $\psi=\bar\psi$, $\Pi=\bar\Pi$,
and $\Phi_i=\partial_i\bar\psi$.  In this limit, no elliptic equation
has to be solved, and the evolution of the field $\bar\Phi_i$ is
effectively ignored when constructing the projection.  We expect that
the optimal choice of $\Lambda$ will be one for which $1/\Lambda$
corresponds to some characteristic length or time scale associated
with the particular problem.  We explore in Sec.~\ref{s:Optimizing}
the properties of these projection operators for a range of $\Lambda$,
and show that an optimal value does exist. When $\gamma_2\neq0$ the
optimal choice seems to be $\Lambda^2=2\gamma_2^2$, where
$1/|\gamma_2|$ is the time scale on which the constraints are
amplified.

Finally, we must consider the boundary conditions for the projection
equations that determine $\psi$, {\it i.e.} Eq.~\eqref{e:helmholtz} or
\eqref{e:poisson}.  In general, boundary conditions for the projection
equations must satisfy two criteria: First, they must be consistent
with boundary conditions imposed on the evolution equations, and
second, the projection equations plus boundary conditions must not
modify solutions that already satisfy the constraints.
Typically, we enforce approximate
outgoing wave boundary conditions on the evolution equations.  For the case of
the scalar wave equation, the approximate outgoing wave boundary
condition, Eq.~(\ref{eq:FreezingBcUm}),
sets $U^{1-}=-\gamma_2\psi$ or equivalently $n^k\Phi_k=\Pi$
on the boundaries (where $n^k$ is the outward directed unit normal).
Since $\Phi_i=\partial_i\psi$ in these projected solutions, the
appropriate boundary condition to impose on $\psi$ in
Eq.~\eqref{e:helmholtz} or \eqref{e:poisson} in this case would be
\begin{equation}
n^k\partial_k\psi=\Pi = \bar\Pi +
\gamma_2(\psi-\bar\psi).\label{e:boundaryeq1}
\end{equation}

Alternatively we can derive boundary conditions for $\psi$ from the
requirement that the boundary variations of the Lagrangian vanish.
The divergence terms in Eqs.~(\ref{eq:dL/dpsi}) and~(\ref{eq:dL/dPhi})
imply that
\begin{eqnarray}
0=n_k\lambda^k=n_k\lambda^{ki},
\end{eqnarray}
on the boundaries for the scalar field system.  A short calculation
(using the fact that $n_k$ is proportional to a gradient, and
$\lambda^{ki}$ is antisymmetric) shows that $n_i\partial_k(g^\half
\lambda^{ki})=0$, so we see from Eq.~\eqref{e:Psibar} that the
natural boundary condition is
\begin{eqnarray}
n^k\partial_k\psi=n^k\bar\Phi_k.\label{e:boundaryeq2}
\end{eqnarray}
If the approximate 
outgoing wave boundary condition, $n^k\bar\Phi_k=\bar\Pi$, was
used in the free evolution step, then the natural boundary condition
Eq.~(\ref{e:boundaryeq2}) differs from Eq.~\eqref{e:boundaryeq1} by
the term $\gamma_2(\psi-\bar\psi)$.  For the constraint projections
described in Section~\ref{s:NumericalResults}, we impose the Robin
boundary condition Eq.~(\ref{e:boundaryeq1}) on the solutions of
Eq.~(\ref{e:helmholtz}) at the boundaries where $U^{1-}$ requires a
boundary condition in the evolution step, and
Eq.~(\ref{e:boundaryeq2}) on the solutions at all other boundaries
({\it e.g.} inside an event horizon).  We note that the discrepancy
between the natural and the physical outgoing boundary condition
could be eliminated by adding an appropriate boundary term to the
constraint projection Lagrangian.

\section{Numerical Results}

\label{s:NumericalResults}

We have studied the effectiveness of the optimal constraint projection
methods developed in Secs.~\ref{s:OptimalProjection} and
\ref{s:ScalarFields} for the case of a scalar field propagating on a
fixed black-hole spacetime.  For these simulations we use the
Kerr-Schild form of the Schwarzschild metric as our background
geometry:
\begin{eqnarray}
ds^2 = -dt^2 + {2M\over r}(dt+dr)^2 + dr^2 + r^2 d\Omega^2.
\label{e:KerrSchieldMetric}
\end{eqnarray}
We express all lengths and times associated with these simulations in
units of the mass, $M$, of this black hole.  Our computational domain
consists of a spherical shell extending from $r_{\mathrm{min}}=1.9M$
(just inside the black-hole event horizon) to $r_{\mathrm{max}}=11.9M$.
For initial data we use a constraint satisfying
Gaussian shaped pulse with dipolar angular structure,
\begin{eqnarray}
\psi&=&0,\label{e:datapsi}\\
\Pi&=&Y_{10}(\theta,\varphi) e^{-(r-r_0)^2/w^2},\label{e:datapi}\\
\Phi_i&=&0,\label{e:dataphi}
\end{eqnarray}
with $r_0=5M$ and $w=1M$.  The value of $\Pi$ is about
$2\times10^{-21}$ at the outer boundary of our computational domain,
below the level of double precision roundoff error.

For the remainder of this section we describe briefly the numerical
methods used to solve this problem.  Then in
Sec.~\ref{s:BoundaryConditionEffects} we describe three numerical
simulations designed to explore the effects of boundary conditions on
the evolution of the constraints in these solutions.  In
Sec.~\ref{s:ConstraintProjectionEffects} we describe two additional
numerical simulations that illustrate the effectiveness of constraint
projection in controlling the growth of constraints.  And finally in
Sec.~\ref{s:Optimizing} we explore ways to optimize the use of the
constraint projection method and measure its computational cost.

All numerical computations presented here are performed using a
pseudospectral collocation method. Our numerical methods are
essentially the same as those we have applied to evolution problems
with the Einstein system~\cite{Kidder2000a, Kidder2001, Lindblom2002,
Scheel2002}, with scalar fields~\cite{Scheel2004}, and with the
Maxwell system~\cite{Lindblom2004}.  Given a system of partial
differential equations
\begin{equation} 
\partial_t u^\alpha(\mathbf{x},t) = 
{\cal F}^\alpha[u(\mathbf{x},t),\partial_i u(\mathbf{x},t) ],
\label{diffeq}
\end{equation}
where $u^\alpha$ is a collection of dynamical fields, the solution
$u^\alpha(\mathbf{x},t)$ is expressed as a time-dependent linear
combination of $N$ spatial basis functions $\phi_k(\mathbf{x})$:
\begin{equation}
u^\alpha_N(\mathbf{x},t) = 
        \sum_{k=0}^{N-1}\tilde{u}^\alpha_k(t) \phi_k(\mathbf{x}).
\label{decom}
\end{equation}
We expand each
scalar function ($\psi$ and $\Pi$) and the Cartesian components of
each vector ($\Phi_x$, $\Phi_y$, and $\Phi_z$) in terms of the basis
functions $T_n(\rho)Y_{lm}(\theta,\varphi)$, where $Y_{lm}$ are
spherical harmonics and $T_n(\rho)$ are Chebyshev polynomials with
\begin{equation}
\rho={2r-r_{\mathrm{max}}-r_{\mathrm{min}}\over 
   r_{\mathrm{max}}-r_{\mathrm
min}}.
\end{equation}
We use spherical harmonics with $\ell\leq \ell_{\mathrm{max}}=5$ and a
varying number of Chebyshev polynomials with degrees $N_r\leq81$.
Spatial derivatives are evaluated analytically using the known
derivatives of the basis functions:
\begin{equation}
\partial_i u^\alpha_N(\mathbf{x},t) 
= \sum_{k=0}^{N-1}\tilde{u}^\alpha_k(t)
  \partial_i\phi_k(\mathbf{x}).
\label{decomderiv}
\end{equation}
Associated with the basis functions is a set of $N_c$ collocation
points $\mathbf{x}_i$.  Given spectral coefficients $\tilde
u^\alpha_k(t)$, the function values at the collocation points
$u^\alpha(\mathbf{x}_i,t)$ are computed by Eq.~(\ref{decom}).
Conversely, the spectral coefficients are obtained by the inverse
transform
\begin{equation}
\tilde{u}^\alpha_k(t) = \sum_{i=0}^{N_c-1} w_i u^\alpha_N(\mathbf{x}_i,t)
                       \phi_k(\mathbf{x}_i), 
\label{invdecom}
\end{equation}
where $w_i$ are weights specific to the choice of basis functions and
collocation points; thus it is straightforward to transform between
the spectral coefficients $\tilde{u}^\alpha_k(t)$ and the function
values at the collocation points $u^\alpha_N(\mathbf{x}_i,t)$.  The
partial differential equation, Eq.~(\ref{diffeq}), is now rewritten
using Eqs.~(\ref{decom})--(\ref{invdecom}) as a set of {\it
ordinary\/} differential equations for the function values at the
collocation points,
\begin{equation} 
\partial_t u^\alpha_N(\mathbf{x}_i,t) 
                      = {\cal G}^\alpha_i [u_N(\mathbf{x}_j,t)],
\label{odiffeq}
\end{equation}
where ${\cal G}^\alpha_i$ depends on $u^\alpha_N(\mathbf{x}_j,t)$ for
all $j$.  This system of ordinary differential equations,
Eq.~(\ref{odiffeq}), is integrated in time using a fourth-order
Runge-Kutta algorithm.  Boundary conditions are incorporated into the
right side of Eq.~(\ref{odiffeq}) using the technique of
Bj{\o}rhus~\cite{Bjorhus1995}. The time step is typically chosen to be
about one fifth the distance between the closest collocation points,
which ensures that the Courant condition is well satisfied.  This
small time step is needed to reduce the time discretization error
to the same order of magnitude as the spatial discretization error at
radial resolution $N_r=61$.

Elliptic partial differential equations, Eq.~\eqref{e:helmholtz} or
\eqref{e:poisson}, are solved using similar pseudospectral
collocation methods.  As detailed in Ref.~\cite{Pfeiffer2003}, we
consider a mixed real/spectral expansion of the desired solution
$\psi(\mathbf{x})$:
\begin{equation}
\psi(\rho_n,\theta,\phi)=\sum_{l=0}^{l_\mathrm{max}}\sum_{m=-l}^l
\hat\psi_{lmn}Y_{lm}(\theta,\phi),
\end{equation}
where $\rho_n$ (for $ n=0, \ldots, N_r-1$) are the collocation points
of the Chebyshev expansion in (rescaled) radius $\rho$.  Given a set
of coefficients $\hat\psi_{lmn}$, we can evaluate the residual of
the elliptic equation and the residual of the boundary conditions
using expressions like Eq.~(\ref{decomderiv}); the requirement that
each $Y_{lm}$ component (for $l\le l_\mathrm{max}$) of this residual
vanishes at the radial collocation points results in a system of
algebraic equations for the coefficients $\hat\psi_{lmn}$.  For the
problem considered here these algebraic equations are linear, and
with suitable preconditioning are solved using standard numerical
methods like GMRES.  The elliptic solver is described in detail in
Ref.~\cite{Pfeiffer2003}.

We use no filtering on the radial basis functions, but apply a rather
complicated filtering rule for the angular functions.  When evaluating
the right side of Eq.~(\ref{odiffeq}), we set to zero the
coefficients of the terms with $\ell=\ell_{\mathrm{max}}$ in the
expansions of the scalars, $\partial_t\psi$ and $\partial_t\Pi$.  The
vector $\partial_t\Phi_i$ is filtered by transforming its components
to a vector spherical harmonic basis, setting to zero the coefficients
of the terms with $\ell=\ell_{\mathrm{max}}$ in this basis, and then
transforming back to Cartesian components.  The result $\psi$ of each
elliptic solve and the projected $\Pi$
(cf. Eq.~[\ref{eq:Pi-projection}]) are filtered similarly.  The
projected $\Phi_i$ is computed via Eq.~(\ref{eq:Phi-Projection}) from
the filtered $\psi$.  We find no angular instability, such as the one
reported in Ref.~\cite{Lindblom2004}, when we use this filtering
method.  And we find no significant change in our results for this
problem by increasing the value of $\ell_{\mathrm{max}}$ beyond the
value $\ell_{\mathrm{max}}=5$.

\subsection{Testing Boundary Conditions}
\label{s:BoundaryConditionEffects}

\begin{figure} 
\begin{center}
\includegraphics[width=3in]{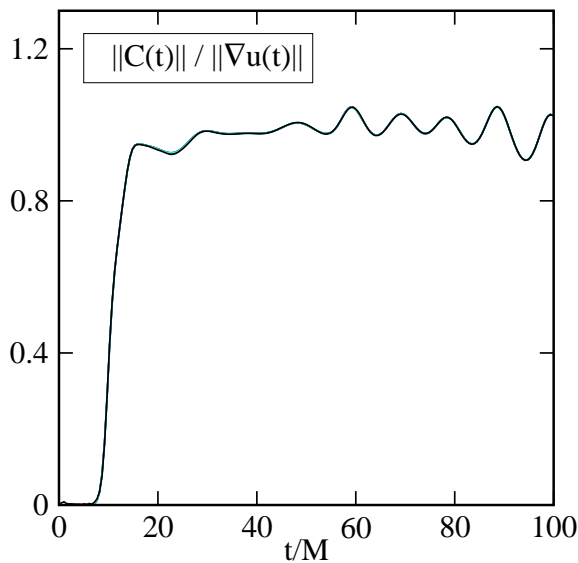}
\end{center}
\caption{Constraint violations for evolutions with
$\gamma_1=\gamma_2=0$, freezing boundary conditions, and no constraint
projections.  Plotted are radial resolutions $N_r=21$, $31$, $\ldots$, $61$;
all curves lie on top of each other.
\label{Fig1}}
\end{figure}

\begin{figure} 
\begin{center}
\includegraphics[width=3in]{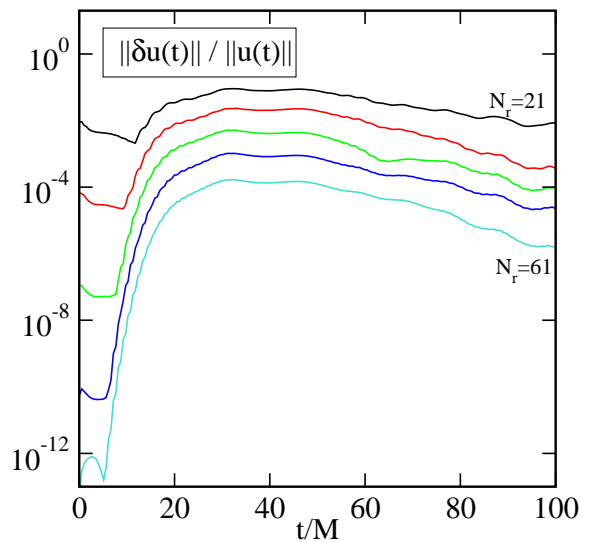}
\end{center}
\caption{Convergence plot for the evolution presented in
Fig.~\ref{Fig1}.  Plotted are differences from the
solution with radial resolution $N_r=81$.
\label{Fig2}}
\end{figure}

In this section we describe the results of three numerical simulations
that explore the effects of boundary conditions on the evolution of
the constraints in the scalar field system.  First we evolve the
initial data in Eqs.~(\ref{e:datapsi})--(\ref{e:dataphi}) using the
standard representation of the scalar field system
($\gamma_1=\gamma_2=0$), and using the standard freezing boundary
conditions on the incoming fields. We use no constraint projection in
this initial simulation.  At the inner boundary of the computational
domain, $r=r_{\mathrm{min}}=1.9M$, all of the fields are outgoing and
so no boundary condition is needed there on any of the fields.  At the
outer boundary, $r=r_{\mathrm{max}}=11.9M$, the fields $Z^1$, $Z^2_i$
and $U^{1-}$ are all incoming since the shift points out of the
computational domain there: $n_kN^k=2M/r$.  So we impose the freezing
boundary conditions $0=\partial_t Z^1=\partial_t Z^2_i=\partial_t
U^{1-}$ on these fields.  The results of this first numerical
simulation are depicted in Figs.~\ref{Fig1} and
\ref{Fig2}.

Figure~\ref{Fig1} illustrates the evolution of the
constraints, which we measure using the quantity
$||C(t)||$,
\begin{eqnarray}
||C(t)||^{\,2}=\int \left(C_iC^i+C_{ij}C^{ij}\right)g^{\half}\,d^{\,3}x,
\end{eqnarray}
divided by a suitable normalization.  The constraints in this system
are combinations of the derivatives of the dynamical fields.  So we
normalize the curves in this figure by the quantity $||\nabla u(t)||$,
which is the natural coordinate-invariant $L^2$ measure of the derivatives
of the dynamical fields:
\begin{equation}
||\nabla u(t)||^{\,2}
=\int  g^{ij}
\nabla_iu^\alpha\nabla_j u^\beta S_{\alpha\beta}\,g^\half\,d^{\,3}x.
\end{equation}
The ratio of these quantities, $||C(t)||/||\nabla u(t)||$, is
therefore a meaningful dimensionless measure of the magnitude of
constraint violations.  When the value of this ratio becomes of order
unity, the dynamical fields do not satisfy the constraints at all.  As
we can see in Fig.~\ref{Fig1}, the constraint satisfying
initial data quickly evolve to a state in which this constraint
measure is of order unity.  A large increase in constraint violation
occurs as the outgoing scalar wave pulse passes through the outer
boundary of the computational domain.  After this time the numerical
solution to the first-order scalar wave system no longer represents a
solution to the original scalar field equation.  

In Fig.~\ref{Fig2} we demonstrate that these numerical
solutions are nevertheless numerically convergent.  We measure the
convergence of these solutions by depicting the quantity
\begin{equation}
||\delta u(t)||^{\,2}= \int S_{\alpha\beta}\left(u_{N_r}^\alpha - u^\alpha_R\right)
\left(u_{N_r}^\beta - u^\beta_R\right)g^\half\,d^{\,3}x,
\end{equation}
divided by a suitable normalization.  This quantity measures the
difference between the solution $u^\alpha_{N_r}$ obtained with radial
resolution $N_r$, compared to a reference solution $u^\alpha_R$.  In
Fig.~\ref{Fig2} we use the numerical solution computed with the
largest number of radial basis functions ($N_r=81$ in this case) as the
reference solution.  In order to make these difference measures
meaningful, we normalize them by dividing by an analogous measure of
the solution itself:
\begin{equation}
||u(t)||^{\,2}= \int S_{\alpha\beta}u_{N_r}^\alpha 
u_{N_r}^\beta\, g^\half\,d^{\,3}x.
\end{equation}
Figure~\ref{Fig2} shows that our computational methods are numerically
convergent, even if the solutions are constraint violating and are
therefore unphysical.  The rate of convergence of these solutions
changes at about $t=10M$ because a short wavelength reflected pulse
enters the computational domain at about this time.  The convergence
of these solutions shows that these constraint violations are a
feature of the evolution system and the boundary conditions, rather
than being artifacts of a poor numerical technique.
\begin{figure} 
\begin{center}
\includegraphics[width=3in]{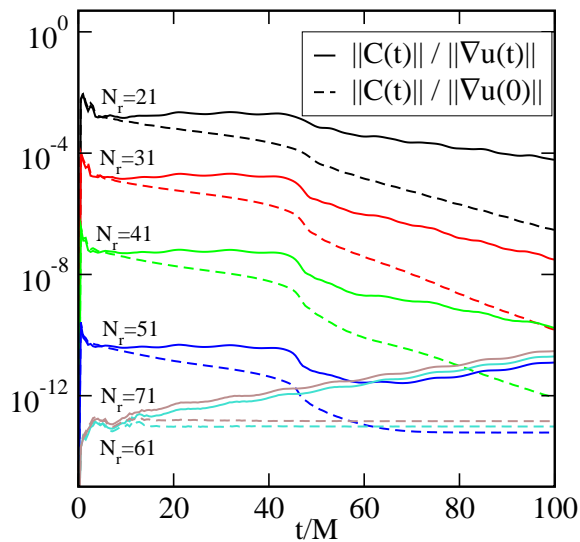}
\end{center}
\caption{Constraint violations for evolutions with
$\gamma_1=\gamma_2=0$, constraint preserving boundary conditions, and
no constraint projection.  Solid curves are normalized by the quantity
$||\nabla u(t)||$ while the dashed curves are normalized by $||\nabla
u(0)||$. Decay of the normalization factor $||\nabla u(t)||$ rather
than growth of the constraints causes the growth in the highest-resolution
solid curves, which have constant roundoff-level constraint violations.
\label{Fig3}}
\end{figure}
\begin{figure} 
\begin{center}
\includegraphics[width=3in]{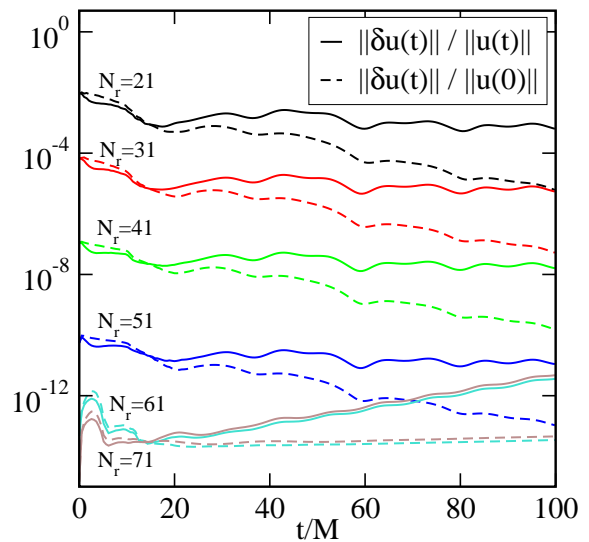}
\end{center}
\caption{Convergence of evolutions shown in
Fig.~\ref{Fig3}.  Plotted are differences from the
evolution with $N=81$, which is henceforth the reference solution
$u_R$.  Solid curves are normalized by $||u(t)||$ while the dashed
curves are normalized by $||u(0)||$.  Decay of the normalization
factor $||u(t)||$ causes the growth in the highest-resolution
solid curves, for which $||\delta u(t)||$ is constant at roundoff level.
\label{Fig4}}
\end{figure}

Next we evolve the same initial data,
Eqs.~(\ref{e:datapsi})--(\ref{e:dataphi}), using the same standard
scalar wave evolution equations ($\gamma_1=\gamma_2=0$), but this time
we use constraint preserving boundary conditions on the fields $Z^1$
and $Z^2_i$, Eqs.~(\ref{eq:ConstraintBcZ1}) and
(\ref{eq:ConstraintBcZ2}).  We use no constraint projection in these
evolutions.  Figure~\ref{Fig3} shows that the
constraints are in fact satisfied by these solutions to truncation
level errors.  The solid curves in Fig.~\ref{Fig3} show
the ratio $||C(t)||/||\nabla u(t)||$ while the dashed curves show
$||C(t)||/||\nabla u(0)||$.  The only difference is that the
denominator used for the dashed curves is time independent.  The solid
curves show that the relative constraint error is approximately
constant in time until about $t=40$, at which time a truncation error
level constraint-violating pulse from the outer boundary has advected
inward across the grid and fallen into the black hole. After $t=40$
the relative constraint error decreases with time. The
highest-resolution solid curves behave differently: they increase
exponentially with time. However, this growth occurs only because the
normalization factor in the denominator (which measures the size of
the derivatives of the fields) goes to zero as the scalar wave pulse
leaves the computational domain.  The highest resolution dashed curves
show that the absolute constraint error for these resolutions
is constant at roundoff level.

Figure~\ref{Fig4} illustrates the numerical convergence of
these evolutions.  Plotted are the ratios of the differences $||\delta
u(t)||$ to a measure of the size of the fields.  The solid curves in
Fig.~\ref{Fig4} show the ratio $||\delta u(t)||/||u(t)||$ while
the dashed curves show $||\delta u(t)||/||u(0)||$.  Again, the only
difference is that the denominator used for the dashed curves is time
independent.  Figures~\ref{Fig3} and \ref{Fig4}
show that these scalar field evolutions are stable, constraint
preserving and numerically convergent.  These solutions therefore
represent what we expect to be the correct physical solution to this
problem.  Were this our only objective, this paper would end here.
However our primary interest here is to study the use of projection
methods to control the growth of constraints.  So we will use the
solution found here as a reference to which our later evolutions using
constraint projection can be compared.

\begin{figure} 
\begin{center}
\includegraphics[width=3in]{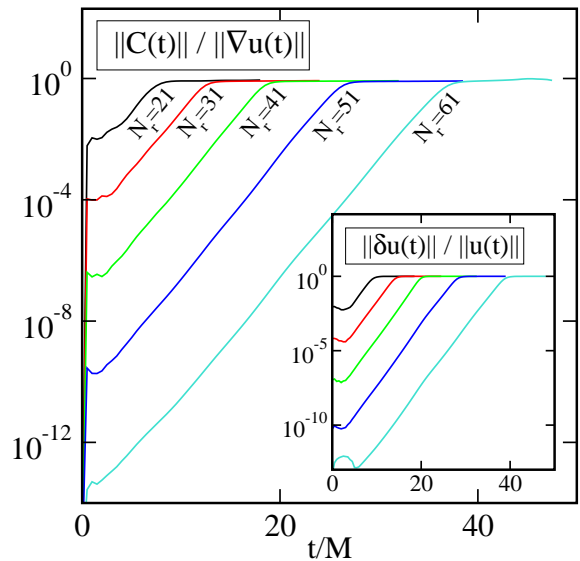}
\end{center}
\caption{Constraint violations for evolutions with $\gamma_2=-1/M$,
constraint preserving boundary conditions, and without constraint
projection.  The inset shows differences $||\delta u(t)||/||u(t)||$ from
the reference solution of Fig.~\ref{Fig4}. The curves level off at
late times because both numerator and denominator grow exponentially
at the same rates.
\label{Fig5}}
\end{figure}

Our last simulation to study the effects of boundary conditions on the
growth of the constraints uses a non-standard scalar field evolution
system with $\gamma_1=0$ and $\gamma_2=-1/M$.  In other respects,
however, this simulation is identical to the one depicted in
Figs.~\ref{Fig3} and \ref{Fig4}: It uses the same initial data,
Eqs.~(\ref{e:datapsi})--(\ref{e:dataphi}), the same constraint
preserving boundary conditions, and no constraint projection.  Because
we use Eq.~(\ref{eq:FreezingBcUm}) as a boundary condition on
$U^{1-}$, the {\it constraint-preserving\/} solutions of the equations
are the same as those obtained with $\gamma_1=\gamma_2=0$.  However,
using an evolution system with $\gamma_2=-1/M$ introduces unstable
bulk terms into the constraint evolution equations,
Eqs.~(\ref{e:const1l}) and (\ref{e:const2l}), so the {\it
constraint-violating\/} solutions of the equations will be different.
Consequently this system is much more pathological than the standard
scalar field system, and provides a much more difficult challenge for
the constraint control methods studied here.  Figure~\ref{Fig5} shows
the evolution of the constraints in this system.  Truncation level
constraint violations in the initial data grow exponentially with an
e-folding time of approximately $1.1M$ in these evolutions. The ratio
$||C(t)||/||\nabla u(t)||$ approaches a constant of order unity at
late times once the constraint violating portion of the solution
dominates and the denominator begins to grow exponentially as well.
The small inset graph in Fig.~\ref{Fig5} illustrates that the
divergence of these solutions from the reference solution of
Fig.~\ref{Fig4} grows at the same rate for all spatial resolutions.
This suggests that the growth is caused by a constraint violating
solution to the evolution equations rather than a numerical
instability.

These evolutions with $\gamma_2=-1/M$ demonstrate that constraint
preserving boundary conditions alone are insufficient to control the
growth of constraints in this system.  Since the Einstein evolution
system is also believed to contain bulk generated constraint
violations~\cite{Lindblom2002}, this example suggests that constraint
preserving boundary conditions alone will not be sufficient to control
the growth of the constraints in the Einstein system.
%
\subsection{Testing Constraint Projection}
\label{s:ConstraintProjectionEffects}

\begin{figure} 
\begin{center}
\includegraphics[width=3in]{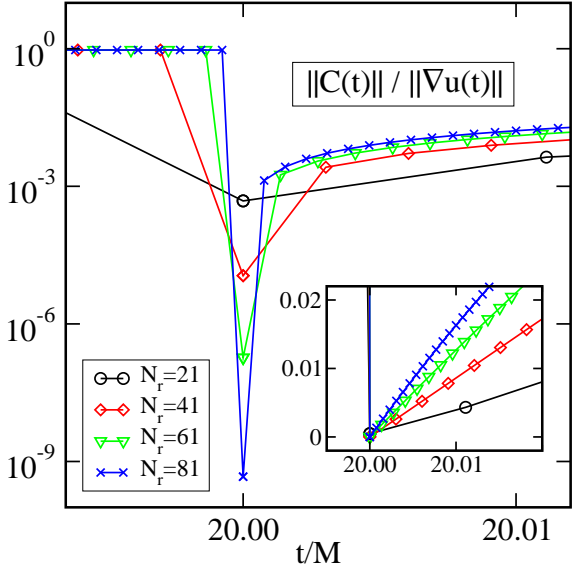}
\end{center}
\caption{Constraint violations for evolutions with
$\gamma_1=\gamma_2=0$, freezing boundary conditions, and a single
constraint projection at $t=20M$ (with $\Lambda=2/M$).  Points show
$||C(t)||/||\nabla u(t)||$ after each time step.  The inset plots the
same data on a linear scale.
\label{Fig6}}
\end{figure}

In this section we discuss two numerical evolutions that explore the
use of the constraint projection methods developed in
Secs.~\ref{s:OptimalProjection} and
\ref{s:OptimalConstraintProjection}.  The first evolution uses the
standard scalar wave evolution system with $\gamma_1=\gamma_2=0$, and
freezing boundary conditions.  We have already seen in
Figs.~\ref{Fig1} and \ref{Fig2} that such
evolutions exhibit significant constraint violations once the scalar
wave pulse passes through the outer boundary of the computational
domain.  In this numerical experiment we freely evolve the scalar
field to the time $t=20M$, and then perform a single constraint
projection on the solution using
Eqs.~(\ref{e:helmholtz})--(\ref{eq:Phi-Projection}) with
$\Lambda=2/M$. We then evolve the system freely again to $t=40M$.
Figure~\ref{Fig6} shows how the constraints respond to
a single constraint projection.  We use a very fine time scale in
Fig.~\ref{Fig6}, showing in detail the times around
$t=20M$ when the constraint projection is performed.  Individual
points in Fig.~\ref{Fig6} show the amount of constraint
violation after each individual time step.  The value of the
constraints drops sharply at the time step where the constraint
projection is performed, and as we expect, the value of the
constraints after this projection step is smaller for higher
resolutions.  So the constraint projection step is successful in
significantly reducing the size of the constraints.  But something
rather unexpected happens next: the constraints increase by orders of
magnitude on the very next free evolution time step after the
constraint projection.  The small inset in
Fig.~\ref{Fig6} shows the same data plotted on a linear
rather than a logarithmic scale.  This shows that the constraints grow
linearly in time after the constraint projection step on a very short
time scale.
%

\begin{figure} 
\begin{center}
\includegraphics[width=3in]{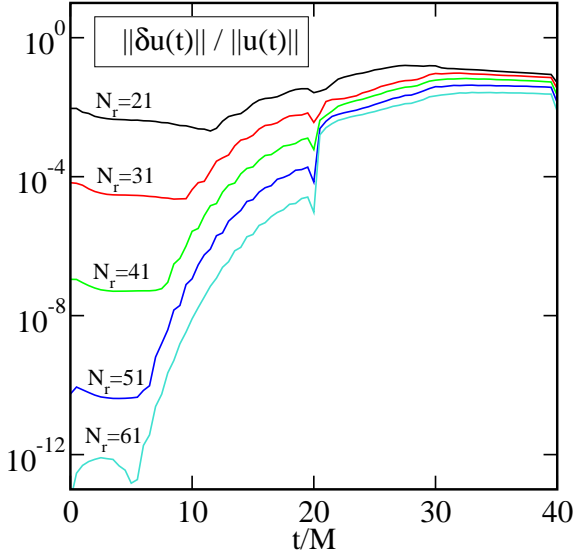}
\end{center}
\caption{Convergence of evolutions shown in
 Fig.~\ref{Fig6}.  Plotted are differences from the
 evolution with $N_r=81$.
\label{Fig7}}
\end{figure}

Figure~\ref{Fig7} provides some information about the reason
for this strange behavior by showing the convergence of these
numerical solutions.  For times before the constraint projection step
at $t=20M$, the solutions show good numerical convergence
as the number of radial collocation points is increased.  But there is a
sharp breakdown of numerical convergence (or at least a sharp drop in
the rate of numerical convergence) after the constraint projection
step.

\begin{figure} 
\begin{center}
\includegraphics[width=3in]{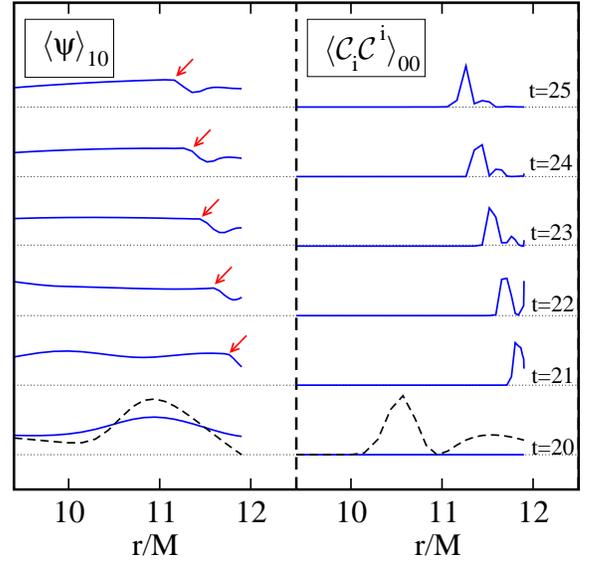}
\end{center}
\caption{Radial profiles of $\langle\psi\rangle_{10}$ and $\langle
C_iC^i\rangle_{00}$ for the evolution of
Fig.~\ref{Fig6}.  The solid lines represent times
$t/M=20, \ldots, 25$.  The dashed line represents the state just
before the constraint projection at $t/M=20$.  The arrows indicate the
location of the non-smoothness in $\psi$.
\label{Fig8}}
\end{figure}

Figure~\ref{Fig8} provides some deeper insight into the reason for
this lack of convergence.  Plotted in Fig.~\ref{Fig8} are a sequence
of curves showing the radial dependences of the dipole part of the
scalar field $\langle \psi\rangle_{10}$ and the monopole part of the
constraints $\langle{\cal C}_i{\cal C}^i\rangle_{00}$ at a sequence of
times including the constraint projection step.  The spherical
harmonic components of a function $Q$ are defined by
\begin{eqnarray}
\langle Q\rangle_{lm} = \int 
Y_{lm}^*(\theta,\varphi)Q(r,\theta,\varphi)\sin\theta\,d\theta d\varphi.
\end{eqnarray}
The dashed lines at the bottom of Fig.~\ref{Fig8} shows the radial
profiles at $t=20M$ immediately before the constraint projection,
while the lowest solid lines show these profiles at the same time
$t=20M$ just after the projection.  We see that the constraints
essentially vanish after the constraint projection step.  The next
profile at $t=21M$ shows that the scalar field develops some
non-smooth radial structure immediately after the projection step,
which subsequently propagates into the computational domain.  This
non-smoothness in $\psi$ causes a sharp spike in the constraints, seen
clearly in Fig.~\ref{Fig8}.  Spectral methods do not converge well
for non-smooth functions, so the emergence of this structure in $\psi$
explains the breakdown in the numerical convergence and thence the
breakdown in our constraint projection method.  The emergence of the
non-smoothness in $\psi$ seems to be caused by the constraint
projection step in the following way: The projection produces a $\psi$
that is non-vanishing at the boundary, and the freezing boundary
condition then forces $\psi=Z^{1}$ (and $Z^2_i$) to develop kinks (see
Ref.\cite{Lindblom2004}) which propagate into the computational domain
during the free evolution steps following the projection.

\begin{figure} 
\begin{center}
\includegraphics[width=3in]{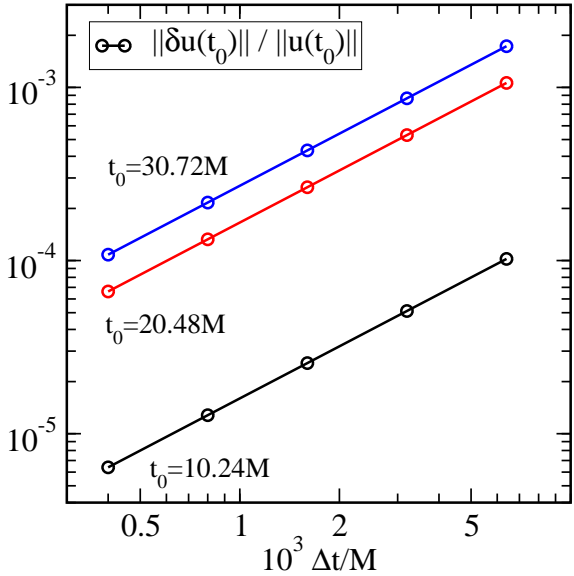}
\end{center}
\caption{Differences between evolutions with time step $\Delta t$ and
the reference solution $u_R$ (of Fig.~\ref{Fig4}) at fixed
evolution times $t_0$. Evolutions use $\gamma_1=\gamma_2=0$, freezing
boundary conditions, and constraint projection with $\Lambda=2/M$ after
each time step, $\DeltaT=\Delta t$.
\label{Fig9}}
\end{figure}

Figures~\ref{Fig6}--\ref{Fig8} demonstrate that
constraint projection is not successful in removing large constraint
violations when used in conjunction with freezing boundary conditions.
One might hope that this failure could be corrected by projecting out
the constraints before they are allowed to grow too large.
Figure~\ref{Fig9} shows the convergence of solutions in which
a constraint projection is performed after each evolution time step,
for a variety of different time steps $\Delta t$.     
Like the evolutions shown in Figures~\ref{Fig6}--\ref{Fig8},
these evolutions
use the standard scalar field system ($\gamma_1=\gamma_2=0$), freezing
boundary conditions, and constraint projection with $\Lambda=2/M$.
The three curves in Fig.~\ref{Fig9} measure the convergence
of the solution (relative to the highest resolution reference solution
depicted in Fig.~\ref{Fig4}) at three different times in this
evolution, $t_0=10.24M$, $20.48M$, and $30.72M$.  All of these
evolutions use the same spatial resolution, $N_r=51$.  These graphs
show that the convergence towards the reference solution is only first
order in the time step $\Delta t$.  This convergence is significantly
worse than that expected for the fourth-order Runge-Kutta time step
integrator that we use.  In contrast the free evolutions with
constraint preserving boundary conditions shown in
Fig.~\ref{Fig4} achieve $||\delta u(t_0)||/||u(t_0)||\lesssim
10^{-10}$ with a timestep similar to the largest $\Delta t$ shown in
Fig.~\ref{Fig9}.  We conclude that constraint projection
produces only first-order in time convergent numerical solutions when
used in conjunction with standard freezing boundary conditions, and
is therefore an ineffective substitute for constraint preserving
boundary conditions.

\begin{figure} 
\begin{center}
\includegraphics[width=3in]{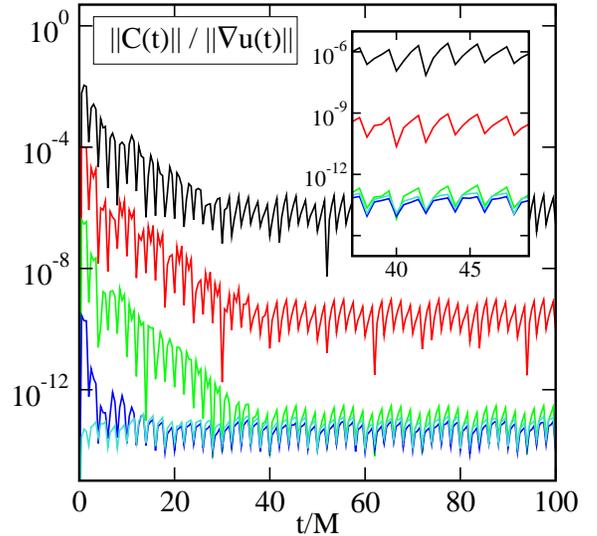}
\end{center}
\caption{ Constraint violations $||C(t)||/||\nabla u(t)||$ for
evolutions with $\gamma_1=0$ and $\gamma_2=-1/M$, constraint preserving
boundary conditions, and constraint projection with $\Lambda=\sqrt{2}/M$ every
$\DeltaT=2M$.  Inset shows the same data with finer time resolution.
\label{Fig10}}
\end{figure}
\begin{figure} 
\begin{center}
\includegraphics[width=3in]{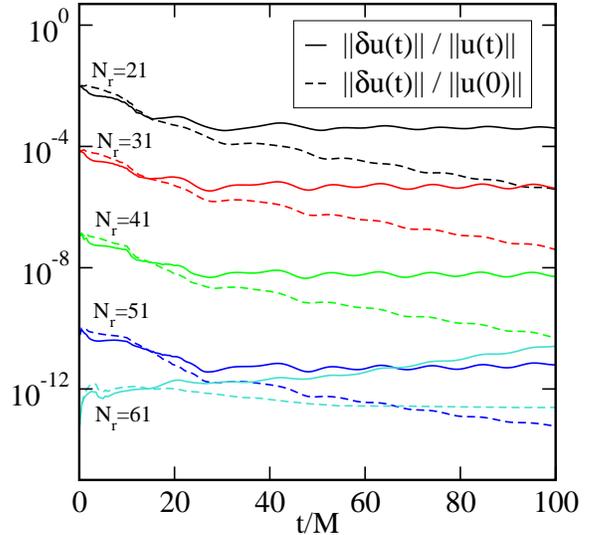}
\end{center}
\caption{Differences from the reference solution $u_R$ (of
Fig.~\ref{Fig4}) for the evolutions shown in
Fig.~\ref{Fig10}.
\label{Fig11}}
\end{figure}
Finally we apply constraint projection to the pathological scalar wave
evolution system ($\gamma_1=0$ and $\gamma_2=-1/M$), which we failed
to control with constraint preserving boundary conditions alone.  We
project every $\Delta T=2M$ using $\Lambda=\sqrt{2}/M$, and we
continue to use constraint preserving boundary conditions.  Except for
constraint projection, this is the same as the evolution shown in
Fig.~\ref{Fig5}.  Figure~\ref{Fig10} shows
that the constraints are reduced to truncation error levels in these
evolutions.  The small inset graph shows these same curves with a
finer time resolution, so the saw-tooth shaped evolution of
the constraints can be seen more clearly.  We note that constraint
projection does not occur at every evolution time step in these
simulations, but rather at fixed times separated by $\DeltaT=2M$.  The
evolutions with the finest spatial resolution take more than one thousand
time steps between projections.  Figure~\ref{Fig11} shows the
convergence between these evolutions and the highest resolution
reference solution depicted in Fig.~\ref{Fig4}.  This figure
demonstrates that the constraint projection method combined with
constraint preserving boundary conditions succeeds in producing the
same numerical solution as our reference solution---even for this
pathological scalar field system.
%

\subsection{Optimizing Constraint Projection}
\label{s:Optimizing}

In this section we explore ways to optimize the use of the constraint
projection methods developed in Secs.~\ref{s:OptimalProjection} and
\ref{s:OptimalConstraintProjection}.  In particular we investigate how
important the choice of the parameter $\Lambda$ is to the
effectiveness of the projection, and we determine its optimal value.  We
also vary the time between projection steps, $\DeltaT$, and determine
the optimal rate at which to perform these projections.  Finally we
measure the computational cost of performing a scalar field evolution
with constraint projection, compared to the cost of doing a free
evolution.
%
\begin{figure} 
\begin{center}
\includegraphics[width=3in]{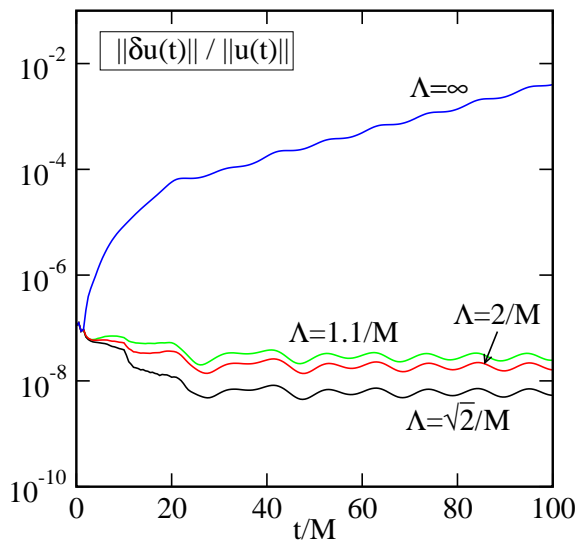}
\end{center}
\caption{Differences $||\delta u(t)||/||u(t)||$ from the reference
solution $u_R$ of Fig.~\ref{Fig4} are plotted for different
choices of $\Lambda$.  Evolutions with $\gamma_1=0$ and $\gamma_2=-1/M$,
constraint preserving boundary conditions, constraint projection every
$\DeltaT=2M$.
\label{Fig12}}
\end{figure}

Figure~\ref{Fig12} shows convergence plots for evolutions of
the pathological scalar field system with $\gamma_1=0$ and
$\gamma_2=-1/M$, constraint preserving boundary conditions, and
constraint projection every $\DeltaT=2M$.  All evolutions use the same
radial resolution, $N_r=41$.  Each of the solid curves in
Fig.~\ref{Fig12} represents an evolution using a different
choice of the parameter $\Lambda$.  We see that the evolutions using
projections with $\Lambda=\sqrt{2}/M$ are somewhat closer to the
reference solution than the others, but the size of the differences
are not very sensitive to the value of $\Lambda$.  The only projected
solution having significantly worse accuracy than the others is the
one with $\Lambda=\infty$, which corresponds to the simple projection
with $\psi=\bar\psi$, $\Pi=\bar\Pi$ and $\Phi_i=\partial_i\psi$.  For
all choices of $\Lambda$, including $\Lambda=\infty$, these evolutions
are exponentially convergent with increasing $N_r$. 

We have some understanding of why there is an optimal choice for the
parameter $\Lambda$:  It is possible to analyze the projection
process completely and analytically for scalar field evolutions with
a flat background metric on a computational domain with three-torus
($T^{\,3}$) topology.  By performing a Fourier transform of the fields in
this case it is easy to show that the fields break up into modes that
propagate with the usual dispersion relation $\omega^2=\vec k\cdot
\vec k$, plus others that grow exponentially in time with dispersion
relation $\omega=i\gamma_2$.  The projection step becomes a simple
algebraic transformation on the Fourier components of the field in
this case.  So it is straightforward to show that the projection step
completely removes the modes that grow exponentially with time only
when the parameters satisfy $\Lambda^2=2\gamma_2^2$.  For evolutions on
computational domains with different topologies, and different
background metrics, it is not possible to determine the optimal choice
of $\Lambda$ using such a simple argument.  However it is not
surprising that the optimal choice is not too different from
$\Lambda^2=2\gamma_2^2$.
\begin{figure} 
\begin{center}
\includegraphics[width=3in]{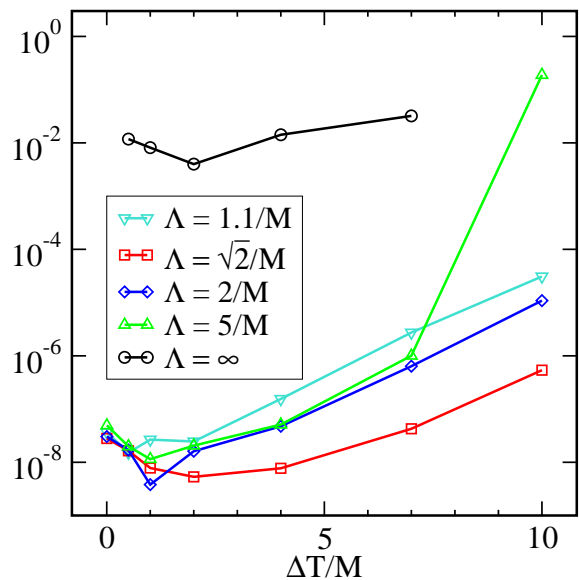}
\end{center}
\caption{Evolutions with $\gamma_1=0$ and $\gamma_2=-1/M$, constraint
preserving boundary conditions and constraint projection every
$\DeltaT$.
Differences from the reference solution $u_R$ (of Fig.~\ref{Fig4})
at $t_0=100M$ for different choices of $\DeltaT$ and $\Lambda$.
\label{Fig13}}
\end{figure}

Next we consider the effect of varying the times between constraint
projections.  Figure~\ref{Fig13} shows the convergence measure
$||\delta u(t_0)||/||u(t_0)||$ for evolutions of the pathological
scalar field system with $\gamma_1=0$ and $\gamma_2=-1/M$,
constraint preserving boundary conditions, and constraint projections
with various values of $\Lambda$ and $\DeltaT$.  These evolutions are
all carried out with the same radial resolution $N_r=41$, and are
compared with the reference solution of Fig.~\ref{Fig4} at the
time $t_0=100M$.  Each curve in Fig.~\ref{Fig13} represents a set
of evolutions with the same value of $\Lambda$ but varying $\DeltaT$.
The smallest $\Delta T$ for each curve corresponds to projecting at
each evolution time step.  We see that all of these curves show a
minimum difference with the reference solution, and this minimum
occurs at about $\DeltaT\approx 1M$ in all of these curves.  This
coincides with the $e$-folding time of the bulk constraint violations,
$-1/\gamma_2$; hence we expect that constraint projection should
generally be applied on a time-scale comparable to that of the
constraint growth.  Figure~\ref{Fig13} also reveals that
projections performed with $\Lambda^2=2\gamma_2^2$ are the optimal
ones over a fairly broad range of projection times $\DeltaT$.
The evolutions with simple constraint projection ($\Lambda=\infty$)
crash for very small values of $\Delta T$, as well as for
$\Delta T=10M$.
\begin{figure} 
\begin{center}
\includegraphics[width=3.4in]{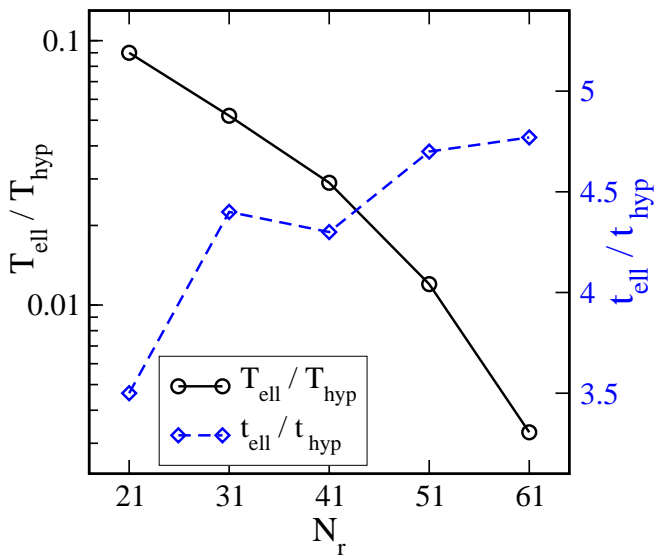}
\end{center}
\caption{Solid curve (left axis) shows the ratio of time spent in
elliptic solves to time spent in the hyperbolic evolution code.
Dashed curve (right axis) shows the ratio of time required for one
elliptic solve to the time for one evolution time step.
\label{Fig14}}
\end{figure}

Finally, we have made some measurements to evaluate the computational
cost of doing scalar field evolutions with constraint projection,
compared to the cost of free evolution.  Figure~\ref{Fig14}
contains two curves that measure the computational cost of doing
optimal projection with $\DeltaT=2M$.  The solid curve shows the
ratio of the time the code spends doing the constraint projection step
({\it i.e.} doing the elliptic solve) $T_{\mathrm{ell}}$ with the time
the code spends doing evolution steps $T_{\mathrm{hyp}}$.  
This ratio decreases from about $0.1$ using a very coarse spatial
resolution to about $0.003$ using a very fine spatial resolution.
The ratio $T_{\mathrm{ell}}/T_{\mathrm{hyp}}$ decreases when the
spatial resolution is increased because the code must take many more
free evolution time steps in the time $\DeltaT$ between projection
steps in this case.  The dashed curve in Fig.~\ref{Fig14}
measures the ratio of the time needed to perform one constraint
projection, $t_{\mathrm{ell}}$, with the time needed to take one free
evolution step, $t_{\mathrm{hyp}}$.  We see that this ratio is fairly
independent of resolution using our spectral elliptic solver, and
ranges from about $3.5$ at low spatial resolution to about $5$ at high
resolution.  These tests show that the computational cost of
performing constraint projection is only a small fraction of the total
computational cost of performing these scalar field evolutions.  We
conclude that computational cost should not be used as an argument
against the use of constraint projection methods.

\section{Discussion}
\label{s:Discussion}

We have developed general methods in Sec.~\ref{s:OptimalProjection}
for constructing optimal projection operators that map the dynamical
fields of hyperbolic evolution systems onto the constraint submanifold
associated with these systems.  These methods are worked out
explicitly in Sec.~\ref{s:ScalarFields} for the case of a new
evolution system that describes the propagation of a scalar field on a
fixed background spacetime.  The constraint projection map for this
system requires the solution of one elliptic partial differential
equation each time a projection is performed.  The new scalar field
system introduced in Sec.~\ref{s:ScalarFields} has the interesting
property that it suffers from constraint violations that flow into the
domain through timelike boundaries and also from violations generated by
bulk terms in the equations.  So this system exhibits both types of
constraint violating pathologies that can occur in the Einstein
evolution system.  To test our constraint projection methods,
we perform a number of numerical evolutions of this
scalar field system propagating on a black-hole spacetime.
We show that constraint
preserving boundary conditions alone are not capable of controlling
the growth of constraints in this scalar field system.  Constraint
projection is also shown to be ineffective when used in conjunction
with traditional boundary conditions that do not prevent the influx of
constraint violations through the boundary.  However we show
that the combination of constraint projection and
constraint preserving boundary conditions is a very effective
method for controlling the
growth of the constraints.  We measure the computational cost of
performing these constraint projections and show that at the highest
numerical resolutions, the projections account for only a fraction of a percent
of the total computational cost of the evolution.  Thus high
computational cost can no longer be cited as a reason to avoid
constraint projection techniques.

\acknowledgments We thank Saul Teukolsky and Manuel Tiglio for helpful
comments. Some of the computations for this project were performed
with the Tungsten cluster at NCSA.  This work was supported in part by
NSF grants PHY-0099568, PHY-0244906 and NASA grants NAG5-10707,
NAG5-12834 at Caltech, NSF grants DMS-9875856, DMS-0208449,
DMS-0112413 at UCSD, and NSF grants PHY-9900672, PHY-0312072 at
Cornell.

\bibstyle{prd} 
\bibliography{References}

\end{document}